\documentclass[a4paper]{spie}  %>>> use this instead for A4 paper
\usepackage{amsmath,amsfonts,amssymb}
\usepackage{graphicx}
\usepackage[colorlinks=true, allcolors=blue]{hyperref}

\usepackage[perpage]{footmisc}
\usepackage{listings}

\title{Italian center for Astronomical Archives publishing solution: modular and distributed}

\author[a]{Marco Molinaro}
\author[a]{Nicola F. Calabria}
\author[a]{Robert Butora}
\author[a]{Sonia Zorba}
\author[a]{Riccardo Smareglia}
\affil[a]{INAF - Osservatorio Astronomico di Trieste, via G.B. Tiepolo 11, Trieste, Italy}

\authorinfo{For further information send correspondence to M. Molinaro - E-mail: marco.molinaro@inaf.it}

\begin{document} 
\maketitle

\begin{abstract}
The Italian center for Astronomical Archives has, among its goals, to provide astronomical data resources as interoperable services based on IVOA standards. It did so for part of its archives (mainly raw telescope data from LBT, TNG and Italian national telescopes) and continued on with hosted data collections and providing expertise to national and international research projects (like WINGS, VIPERS, VIALACTEA).
Its expertise and knowledge of the VO comes from active participation within IVOA and VO at European and international level, with a double-fold goal: learn from the collaboration (acquiring skills and technical knowledge) and provide inputs (implementations and feedback) to the VO community.
In this scenario the first solution to build an easy to configure and maintain resource publisher conformant to VO standards proved to be too optimistic, not considering the complexity the IVOA architecture could have reached in a short while. For this reason it has been necessary to re-think the architecture with a modular system.
This latter is now partially in place and will gradually replace the previous solution allowing for an easier to extend and rework if major changes will happen at the level of VO standards.
The solution chosen for the architecture orbits around the messaging concept, where each modular component speaks to the other interested parties through a system of broker-managed queues (currently using AMQP with RabbitMQ as the broker).
The messaging system lets us free to choose the development language for the business logic components, not only for the front-end, web interfacing solutions (needed to expose VO HTTP based protocols), but also on the archives and database access components, the logging systems and any other tool or component that may be needed in the future.
The first implementation covered the simplest VO protocol, the Simple Cone Search, were the messaging task architecture connects the parametric HTTP interface to the database backend access module, the logging module, and allows multiple cone search resources to be managed together through a configuration manager module. Even if it has been initially used as a test for the new architecture, it already proved the flexibility required by the overall system when the database backend needed to be changed from a MySQL to a PostgreSQL+PgSphere solution.
Another implementation test has been made to leverage task distribution over multiple servers dedicated to computation to allow for a single HTTP interface to serve simultaneously: FITS cubes direct linking, cubes cutout (using an AST and cfitsio engine) and cubes positional merging (using a Montage based solution). The solution proved also to be a quick answer to load distribution (although not really efficient).
Alongside these the implementation of the SIA-2.0 standard protocol is ongoing, following the scheme used for the Simple Cone Search, while for the TAP protocol implementation we will be re-using and adapting the already available TAPlib library.
Alongside this production, message-driven, publishers, a first administration tool (TASMAN) has been developed to ease the build up and maintenance of the TAP\_SCHEMA component of TAP services including also ObsCore maintenance capability.
Future work will be devoted at widening the range of VO protocols covered by the set of available modules, improve the configuration management and develop specific purpose modules common to all the service components.
\end{abstract}

% Include a list of keywords after the abstract 
\keywords{data providing, virtual observatory, micro-servicing, message broker}

\section{INTRODUCTION}
\label{sec:intro}  % \label{} allows reference to this section
The Italian Center for Astronomical Archives\footnote{\url{http://ia2.inaf.it}} (IA2) manages the archives of ground based telescopes, like the Telescopio Nazionale Galileo\footnote{\url{http://www.tng.iac.es/}} (TNG, Canary Islands), the Large Binocular Telescope\footnote{\url{http://www.lbto.org/}} (LBT, Arizona) and the Asiago Observatory\footnote{\url{http://www.oapd.inaf.it/index.php/it/asiago-home.html}} (Italian Alps), provides support to archive and data-providing facilities for projects and simulated data, like WINGS (WIde-field Nearby Galaxy-cluster Survey), VIPERS (VIMOS Public Extragalactic Redshift Survey), VIALACTEA ("The Milky Way as a Star Formation Engine") and provides infrastructure services for the Italian national Institute for Astrophysics\footnote{\url{http://www.inaf.it}} (INAF). IA2 is a project of INAF under its ICT office umbrella and, among its tasks, it tried to follow the Virtual Observatory (VO) community, lead by the International Virtual Observatory Alliance\footnote{\url{http://www.ivoa.net}}, in providing its data resources in an interoperable way. 

To do so initially a web application was devised to take care of the IVOA so called \textit{simple protocols} (ConeSearch [\citenum{2008ivoa.specQ0222P}], SIA [\citenum{2009ivoa.spec.1111H}], SSA [\citenum{2012ivoa.spec.0210T}]). This application, named VO-Dance (see [\citenum{2012SPIE.8451E..05M}], [\citenum{2012ASPC..461..419M}]), originated from the idea that the parametric query interface of the simple protocols would have stayed homogeneous (which proved false with the advent of the Table Access Protocol - TAP, [\citenum{2010ivoa.spec.0327D}]), and tried to work on a simple mapping principle for dataset metadata to cope with the internal archival solution and expose the required metadata and annotations mandated by the IVOA Recommendations.

\begin{figure} [ht]
   \begin{center}
   \begin{tabular}{c} %% tabular useful for creating an array of images 
   \includegraphics[width=.7\textwidth]{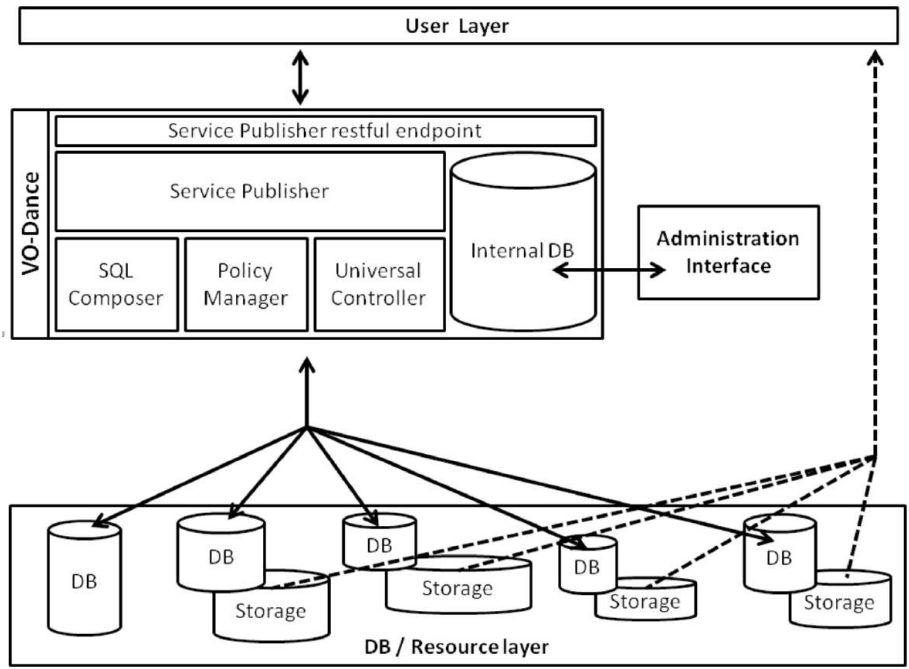}
   \end{tabular}
   \end{center}
   \caption[vodancearch] 
%>>>> use \label inside caption to get Fig. number with \ref{}
   {\label{fig:vodancearch}
Sketchy representation of the VO-Dance architectural blocks. Worth noting the common ReST interface and publisher blocks that served all the services configured for the web application.}
   \end{figure}

The advantages of this approach were a quick and simple configuration management for each single service added to a (tentatively) homogeneous interface. The drawbacks proved to be in the complexity raising in extending it to new protocols and the monolithic approach that bound together the various services, with the consequence that, failing one service, the others usually became unreachable. Figure \ref{fig:vodancearch} depicts the architecture of the VO-Dance web application.

\begin{figure} [ht]
   \begin{center}
   \begin{tabular}{c} %% tabular useful for creating an array of images 
   \includegraphics[width=.7\textwidth]{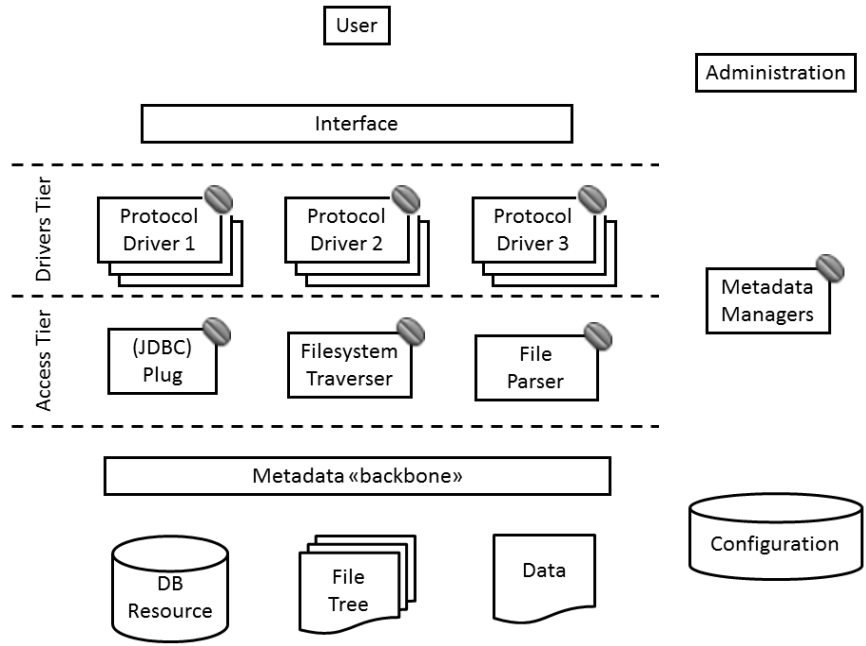}
   \end{tabular}
   \end{center}
   \caption[vodancearch] 
%>>>> use \label inside caption to get Fig. number with \ref{}
   {\label{fig:ejbarch}
Initial, EJB based, solution to overcome the monolithic VO-Dance limitations. The Access tier already shows multiple possible data base back-end access solutions.}
   \end{figure}

To overcome the limitation of VO-Dance it was then investigated the idea of using a modular solution (see [\citenum{2014SPIE.9152E..0CM}]), so that each component could be replaced or rewritten without the need to touch the full infrastructure and also to allow modules to be run in multiple instances, foreseeing load distribution. The first attempt made use mainly of Java technologies (like the EJB, Enterprise Java Beans) and, while allowing for load distribution, still it was lacking the general scope of modularity, e.g. in letting multiple languages to concur to the architecture and requiring explicitly full web application containers in place to work properly. Figure \ref{fig:ejbarch} sketches the proposed EJB modular solution.

Subsequently a more extensive search was made to identify what building blocks were needed to build a modular architecture for resource and service publishing. This was done in the scope of a Degree Thesis in Informatics, (unfortunately in Italian only, Francesco Cepparo\footnote{"Sviluppo di un sistema distribuito modulare e scalabile per la pubblicazione di servizi VO" (Development of a distributed, modular and scalable system for VO service publishing), University of Udine, 2014.}) and reported in [\citenum{2016SPIE.9913E..28C}]; the final solution, which led to the results in this contribution, consists in using a messaging system to build a workflow for the resources' servicing architecture that is deployed as a set of modular services.

This latter architecture structure (see Sec.~\ref{sec:modular}) has then been used to start re-coding the publishing modules. In Ref. [\citenum{2016SPIE.9913E..28C}] already a couple of examples of Simple Cone Search deployments were mentioned (using multiple coding languages and different back end solutions). After that the modules have been revisited (see Sec.~\ref{subsec:scs}) and new modules are nearly ready for production (see Sec.~\ref{subsec:sia}).

A special role in the architecture that will be described in the next section (Sec.~\ref{sec:modular}) is played by the IVOA TAP service. This is the first (but hopefully not only) place where one of the modules is directly taken from an existing implementation of a standard. The specific role played by TAP in the architecture will be explained in Sec.~\ref{subsec:taprole}, that hosts also a sub-section (Sec.~\ref{subsubsec:taplib}) that discusses briefly the external library that is being used.

Besides the architecture description in Sec.~\ref{sec:modular}, an overview of what part of the VO scenario is involved at current status of implementation is reported in the introduction to Sec.~\ref{sec:voimpl}, where the subsections describe the single modules.

Later (Sec.~\ref{sec:vlkb}) we present a specific, custom based, set of resources and services that are designed with the same architecture in mind and will eventually be merged in the IA2 global publishing framework. Sec.~\ref{sec:other} will then briefly describe what other independent modules exist or are foreseen.

The last two sections are devoted to a tentative summary of the advantages and disadvantages of this solution (Sec.~\ref{sec:procon}), as they appear at current stage, and some final conclusions Sec.~\ref{sec:conclusions}).

\section{Modular approach to resource publishing}
\label{sec:modular}

The were various reasons to move from the VO-Dance one to a more modular/flexible approach:
\begin{itemize}
\item abandoning Java Reflections, which served in an homogeneous system of interfacing, but was heavier than expected on performance; moreover the VO scenario evolved in a way to make the reflection not optimal with respect to re-coding specific interface parts;
\item adapting to different types of query interfaces than simple parameter-based ones;
\item moving from a service-oriented scenario to a resource-oriented one, where services play really the role of interfaces to the underlying dataset collections;
\item allowing to work in a distributed environment, with modules that can be deployed multiple times to lower the load on single system components;
\item try to re-use existing software libraries and tools from third parties (and integrate them in the architecture);
\item support for multiple coding languages to allow specialized languages when needed (i.e. not relying on Java only, as it was before).
\end{itemize}

These were the main ones, defined already at the beginning of the process of reshaping the publishing architecture. What was also kept in mind was the initial vision that the architecture should form a second layer on top of an existing archival system, as will be described better in Sec.~\ref{subsec:taprole} and was described in [\citenum{2014ASPC..485..139M}]. Moreover, it was clear that administration tools were required: these are left currently out of the efforts (with one major exception, see Sec.~\ref{subsubsec:tasman}), letting configuration steps to live in manual management of files and processes startup while the framework reaches an higher maturity level.

\begin{figure} [ht]
   \begin{center}
   \begin{tabular}{c} %% tabular useful for creating an array of images 
   \includegraphics[width=.5\textwidth]{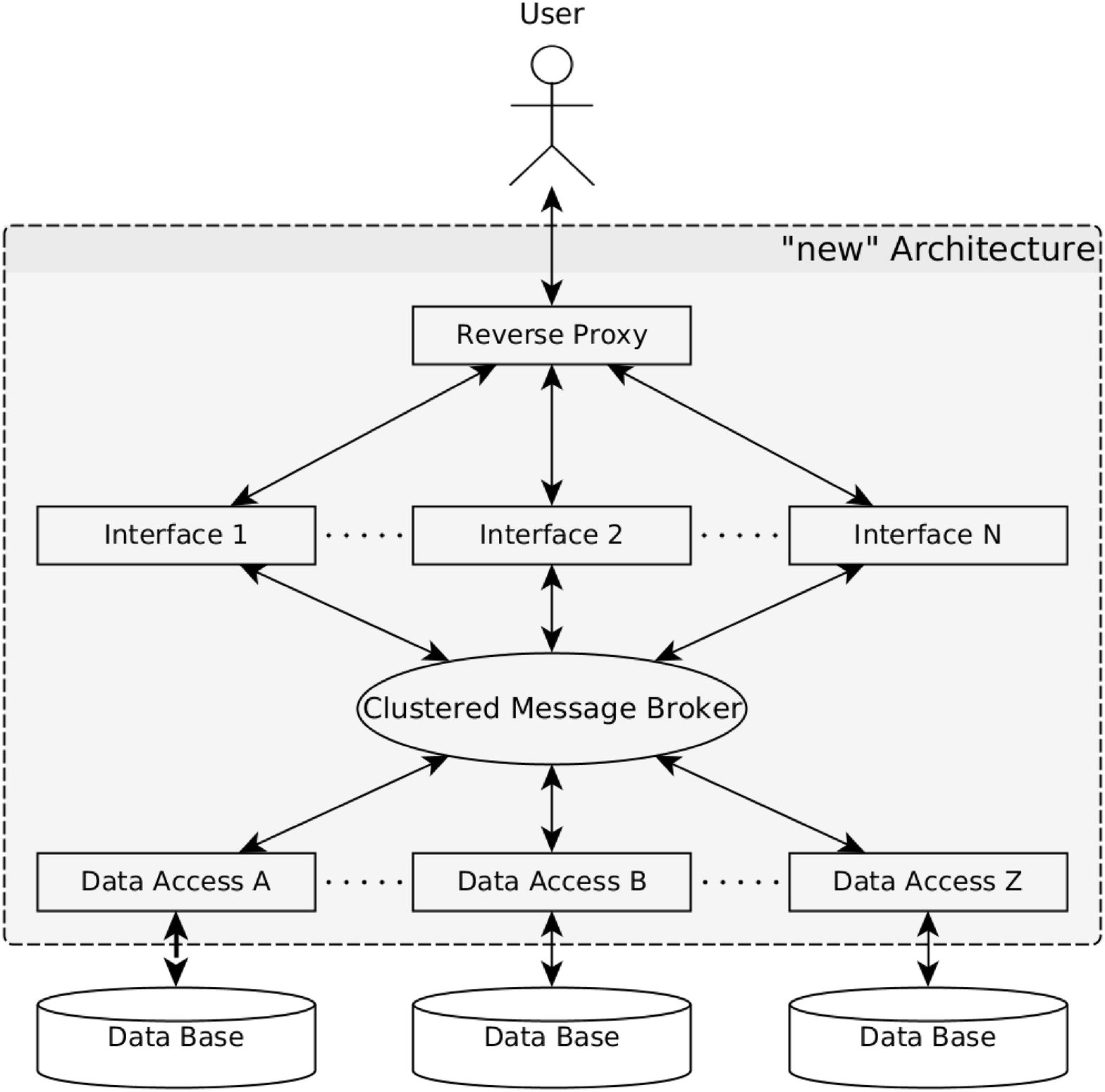}
   \end{tabular}
   \end{center}
   \caption[newarch] 
%>>>> use \label inside caption to get Fig. number with \ref{}
   {\label{fig:newarch}
Sketchy diagram that describes the basic elements of the currently used architecture for publishing VO resources at the INAF IA2 archives center. It shows the use of reverse proxying to distribute the load of the query calls among the query interface protocols, the message brokering solution to create the workflow for the jobs meant to deliver the response out of the incoming request and the layering subdivision among the interface layer modules and the modules explicitly used to perform data access.}
   \end{figure} 

Evaluation of tools and components, during the previously cited degree thesis, brought to a really simple set of modules to be prepared and set in place. Figure~\ref{fig:newarch} sketchily depicts it and shows at least that:
\begin{itemize}
\item a proxy solution is in place to distribute the requests load on the interfaces;
\item a message broker takes care of delivering the right messages among the various modules;
\item a separation is in place (taken care by the broker) between the interfaces accepting the user request and the access modules actually performing the queries.
\end{itemize}

The tools identified to build this system, and currently in place, are:
\begin{itemize}
\item HAProxy\footnote{\url{http://www.haproxy.org/}} as the proxy engine;
\item AMQP\footnote{Advanced Message Queuing Protocol - \url{https://www.amqp.org/}} as the message protocol;
\item RabbitMQ\footnote{\url{https://www.rabbitmq.com/}} as the AMQP broker implementation.
\end{itemize}

Those have being chosen because they are quite simple (HAProxy, AMQP) or because there exist enough implementations to cover message management in various coding languages (AMQP).

\begin{figure} [ht]
   \begin{center}
   \begin{tabular}{c} %% tabular useful for creating an array of images 
   \includegraphics[width=.5\textwidth]{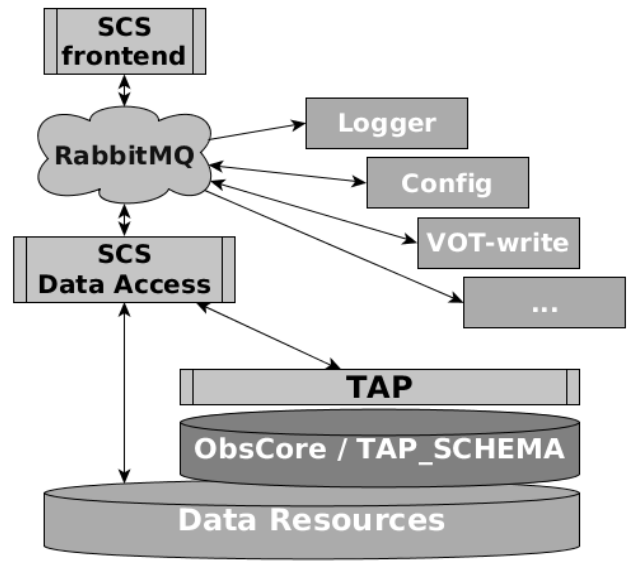}
   \end{tabular}
   \end{center}
   \caption[ia2blocks] 
%>>>> use \label inside caption to get Fig. number with \ref{}
   {\label{fig:ia2blocks}
Graphical representation of the module blocks for the Simple Cone Search implementation. The \textit{VOT-write} is currently not an implemented independent module, but VOTable serialization in included in the data access module. The arrow connecting the vb-cs data access module to the TAP service is not in place, even if it would be easy to mimic a cone search using an ADQL query.}
   \end{figure}

The architecture shown in Fig.~\ref{fig:newarch} actually skips on the other usage for the message broker, i.e. the possibility to connect the request/response flow with ancillary modules dedicated to specific tasks, like: logging, configuration retrieval, output specific serializations. Figure~\ref{fig:ia2blocks} expands on this showing the case of the Simple Cone Search with some additional components (some of them already available, some others not). Anyway, the architecture keeps its basic meaning and, apart from the needs of a VO aware publishing solution, it can be used to accommodate also custom services, like it will be described in Sec.~\ref{sec:vlkb} for the services of the VIALACTEA Knowledge Base (VLKB, see, e.g. [\citenum{2016SPIE.9913E..0HM}]).

A last point to take into account in understanding this solution is the specific role that the TAP protocol plays. For the services implementing this protocol there will be no division of the module interface accepting requests from the data access part and this for a simple reason: the TAP services, and their internal TAP\_SCHEMA components, will work as the metadata container for the other services. More on this in Sec.~\ref{subsec:taprole} and subsections.

\section{Virtual Observatory scenario}
\label{sec:voimpl}

The main reasons to restart building the publishing architecture of the data center were: allowing a (quite) easy deployment of VO aware or compliant resources, and do that while keeping in mind custom requirements from the archives center. It will be shown later how it is possible to cope with the two requirements together (Sec.~\ref{subsec:taprole}) while here we briefly describe the VO scenario this publishing system lives in.

\begin{figure} [ht]
   \begin{center}
   \begin{tabular}{c} %% tabular useful for creating an array of images 
   \includegraphics[width=.5\textwidth]{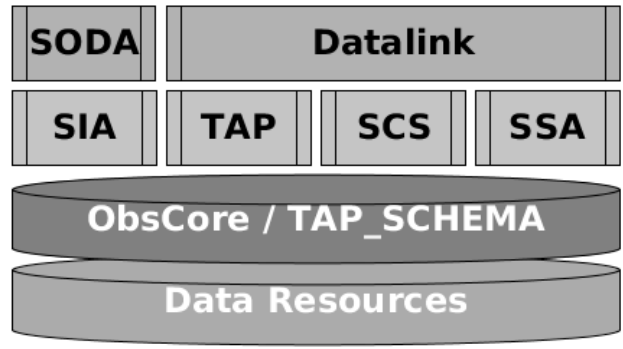}
   \end{tabular}
   \end{center}
   \caption[voblocks] 
%>>>> use \label inside caption to get Fig. number with \ref{}
   {\label{fig:voblocks}
Simplified view of a subset of Data Access Layer services on top of a (generic) data resource. Three layers on top of the data resource are sketched: a metadata content layer (ObsCore/TAP\_SCHEMA), a dataset discovery one (SIA, TAP, SCS and SSA, even if SCS, the Simple Cone Search actually performs data access directly), and a dataset access (SODA and DataLink).}
   \end{figure}

The IVOA architecture (see [\citenum{2010ivoa.rept.1123A}], diagrams therein) describes the \textit{Data Access Protocols} as the connection between users and providers that allows the users to access the data resources on the providers side. There are many access protocols but we can try to simplify the view on how current ones (most of) can work together from Fig.~\ref{fig:voblocks}.

On top of a data resource (where, by resource, we mean generically a collection of datasets/files and the basic metadata to reach them - location, size, \ldots - maybe stored into an RDBMS) VO specifications define, through models and semantic annotations, specific metadata meant to make interoperable the resource they describe.

In the case of using a TAP service, tabular data can have these annotations stored in the TAP\_SCHEMA component of the service; if data is not directly in tabular form, i.e. we have a set of files (collection of datasets) the metadata annotation for each dataset might be stored in a special table, the \textit{ivoa.obscore} one (see [\citenum{2017ivoa.spec.0509L}] for details on the IVOA ObsCore Recommendation), that, in turn, can be used by TAP or SIA version 2.0 (see [\citenum{2015ivoa.spec.1223D}]) for discovery purposes.

Once these annotations are in place, and define the collections of data to deploy, a set of discovery services can sit on top of them. Table Access Protocol services, or Simple Cone Searches can live on top of catalogs or tabular data; image or spectra datasets discovery (through TAP, SIA or SSA [\citenum{2012ivoa.spec.0210T}]) can be used to identify the data to be later accessed via direct download of performing specific operations server-side on top of them.

This latter part is currently standardized in the SODA (Server-side Operation for Data Access [\citenum{2017ivoa.spec.0517B}]) and DataLink [\citenum{2015ivoa.spec.0617D}] protocols and allow direct, standardized or custom access to the datasets identified in the discovery response phase outlined above.

From this simplified view and description one can understand why a modular approach can be useful, with various layers applied from the actual data up to the user's access (not to consider that the above described standards are attached to other ones, see later in Sections from~\ref{subsec:scs} to~\ref{subsec:taprole} or~\ref{sec:other} for additional details).

Another requirement coming from the VO scenario, and directly impacting the access layer specifications above, is the use of spherical geometry in data discovery and access requests. This request is clear from specifications like Cone Search and SIA, and impacts TAP services because the only mandatory query language for this protocol is the ADQL (Astronomical Data Query Language [\citenum{2008ivoa.spec.1030O}]). ADQL defines geometrical types and functions to be used in tabular queries. The way this specification impact the metadata layer preparation and the effort needed from an RDBMS back-end will be expressed in the next subsections.

Out of the above scenario and requirements, considering the approach chosen and introduced in Sec.~\ref{sec:modular}, a work has been started to build module blocks to implement VO resources deployment. Hereafter we will describe current status of modules with some hints on the underlying resource back-end, but not fully describing the data resources and collection preparation that lies behind all of these.

\subsection{Simple Cone Search implementation}
\label{subsec:scs}

The first protocol to be implemented has been the Simple Cone Search. This was an easy choice given the really simple query request interface it presents and the minimal basic metadata requirements it asks for.

The modules needed to run a service are currently grouped under the \textit{VOBall} project and consist of:
\begin{itemize}
\item a web module (vb-web): to accept the user's HTTP requests;
\item a cone server module (vb-cs): to handle the query on the chosen back-end;
\item a configuration repository module (cfg-repo): to manage configurations for the services;
\item a logging module.
\end{itemize}

All of the above are connected through AMQP messages exchanged through a RabbitMQ broker. The web module is a Java web application that requires a GlassFish web application container. The other three modules are Java applications. The messages that the modules exchange are currently JSON objects that include, after the recipient module identification, the message needed to move on the next step in the request/response workflow.

With the currently developed modules up to three (potentially different) RDB resources are required:
\begin{description}
\item [data source] the vb-cs module connects to it to retrieve the catalog part identified by the query request constraints. Current module is able to speak to MySQL, PostgreSQL and PostgreSQL augmented by the PgSphere\footnote{\url{https://github.com/akorotkov/pgsphere}} plugin.
\item [metadata source:] VOBall attaches itself to a TAP\_SCHEMA to annotate responses properly and operate on the right query filters. This TAP\_SCHEMA can be stored in PostgreSQL and MySQL instances.
\item [logging:] the log module is currently limited to a MySQL back-end.
\end{description}

The PgSphere solution for the data source allows for an easier circular positional cut on the deployed catalog resource, using the \textit{spoint} PgSphere added data type, while simple MySQL and PostgreSQL back-end require trigonometrical functions into the query, thus slowing down the SQL response.

The metadata source stored into a TAP\_SCHEMA schema are meant to allow VO aware metadata annotation for the deployed catalog fields. Of course, different solutions may be chosen, but this one allows for annotating the data source in one place and then possibly re-use the annotation in other service capabilities. Note that, if one wants only Cone Search, this requirement on the TAP\_SCHEMA does not mean having a full TAP service in place, but only its metadata content aligned to that specification.

As for the logging DB, the module currently is bound to use MySQL but changes in this module are already foreseen in the near future (see also Sec.~\ref{sec:other}) and, since this is really meant to be a modular solution, the logging module could be easily replaced, also because it is currently really a bare implementation.

To complete the working architecture alongside software modules, AMQP broker and RDB back-ends, what is missing are the configurations for the various services.
The set of services above rely on a set of configuration files, stored inside a subfolder of the home directory of the user deploying the services. These configuration files are read by the cfg-repo module and accessed by the other modules at startup through calls to the cfg-repo itself.

There are three configuration types to fill in:
\begin{description}
\item [TAP\_SCHEMA connectors] that contain connection information to the - possibly various - TAP\_SCHEMA metadata sources in use by the full stack of cone services. 
\item [Cone Service configurations,] one for each deployed service, storing the key=value pairs identifying and describing the service from the operational point of view, including settings to contact the broker, retrieve metadata, send logs, \ldots
\item [Logging connectors] for configuring log repo connection(s).
\end{description}

All of these are simple \textit{\{name\}.properties} textual files where \{name\} is the identifier for the connectors or services to be shared within the configuration system (i.e. if you have a \textit{foo\_tap\_schema} connector it will have its configuration stored into a \textit{foo\_tap\_schema.properties} file and it will be referenced inside a \textit{myservice.properties} file using a \textit{tapServiceName} =\textit{foo\_tap\_service} key=value pair).

The listing~\ref{lst:props} shows an example for a service configuration file. Within its contents you may see connection details for the data source, messaging and logging configuration and a few service specific details, including the PgSphere provided spoint field used for the circular cut on the catalog content.

\begin{center}       
\begin{tabular}{c} 
\begin{lstlisting}[language=sh,caption={Sample \textit{service.properties} content},captionpos=b,label={lst:props}]
dbUnit=Degrees
dbName=mydb
schemaName=myschema
tableName=mytable
dbUser=user
dbPassword=secret
dbHost=dbserver
dbPort=5432
dbType=PostgreSQL
brokerHost=rabbithost
brokerPort=rabbitport
tapServiceName=mytapschema
loggingServiceName=logserviceid
pgSphereColumn=pointcoords
\end{lstlisting}
\end{tabular}
\end{center}

Alongside single service and connectors configuration two other configuration files are used:
\begin{itemize}
\item one \textit{cone.properties} file to list the actual services to be deployed;
\item one \textit{logging.properties} to list the logging connectors in use.
\end{itemize}

Both are simple comma-separated identifiers list. Services or connectors not listed in those files, even if configured, will not be made available.

The modules in this, or similar, flows have some precedence rule. As it has been said the configuration repository module is asked by web and access ones about what resource they will be deploying. Also the logging service has to be listening before the first logging messages come from the services. For this reason config repository and logging modules have to be up and running before the service modules for access and web. Similarly, all the modules require the broker to be running to be able to exchange messages. Trivial to say that data sources and configurations have to be in place, and that, if multiple modules provide the same service, the proxy solution has to be configured and running as well.

So far, testing and pre-production services have been deployed using this \textit{VOBall} modules set. The modules are an improvement over VO-Dance, also in terms of performance, especially considering the PgSphere geometrical solution in place. This latter change was the first check the the modularity was an help in development and maintenance. With VO-Dance the full application would have been touched by the back-end change, while here only the access module needed to be touched and tested.

\subsection{Simple Image Access v2.0}
\label{subsec:sia}

After successfully moving the Simple Cone Search implementation from the former VO-Dance solution to the modular one, it was decided to start the implementation of a set of modules for the Simple Image Access version 2.0. This latter is actually a protocol for collection and dataset discovery in the general case of multi-dimensional datasets, thus able to expose images, cubes, spectra and other atomic datasets in an homogeneous way, mostly using the same metadata the ObsCore specification uses.

This implementation task is currently ongoing, using the same modules approach and components from the cone search one, only specializing on the image access needs. Since most of the metadata requirements are similar or identical to the ObsCore ones a decision has been taken to rely on the \textit{ivoa.obscore} table to provide the discovery information.

This is not the only requirement in place that makes this modules more specific to a back-end implementation. The requirements on spherical geometry filtering of the datasets collections made it also depend directly on the PostgreSQL and PgSphere database solution.

As for the cone search, where one can think about an access module that simply uses an ADQL query over an underlying TAP service using the TAP\_SCHEMA metadata already in place for the service itself, for the SIA access one could think of a similar module, using an ObsTAP (TAP featuring ObsCore model) service behind the scenes. This could probably be the way also for an SSA module that is not yet foreseen, mostly because SSA and SIA are somehow overlapping in the current VO scenario.

\subsection{Table Access Protocol and TAP\_SCHEMA roles}
\label{subsec:taprole}

\begin{figure} [ht]
   \begin{center}
   \begin{tabular}{c} %% tabular useful for creating an array of images 
   \includegraphics[width=.4\textwidth]{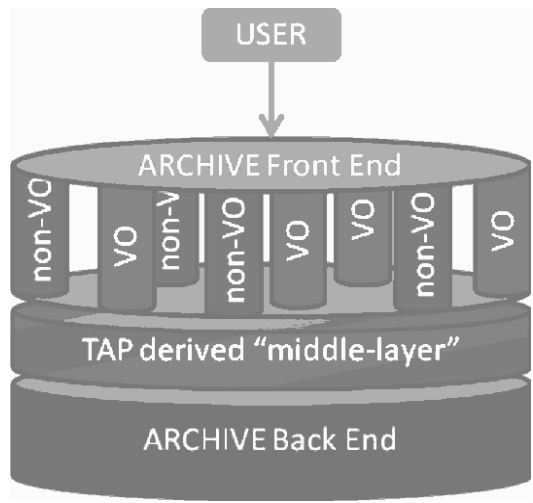}
   \end{tabular}
   \end{center}
   \caption[tapbackbone] 
%>>>> use \label inside caption to get Fig. number with \ref{}
   {\label{fig:tapback}
A set of resources forming the archive beck-end can be enriched by VO driven metadata using the TAP\_SCHEMA defined by the IVOA TAP protocol. Given this schema is extensible in terms of content, this means it is possible to accommodate the custom archive needs alongside the VO ones, thus letting VO standardized and custom services work alongside while accessing the same metadata and data for discovery and access purposes.}
   \end{figure}

As hinted previously, TAP services and their TAP\_SCHEMA metadata components play a special role in this modular architecture. The underlying idea comes from trying to let VO driven resource publishing live together with custom data providers solutions (see, e.g., [\citenum{2014ASPC..485..139M}] from which Fig.~\ref{fig:tapback} has been adapted). The idea is that, given TAP metadata specification allows for custom additions and extensions, then discovery services of different kinds can live on top of it (as sketchily shown in Fig.~\ref{fig:tapback}).

Even if this is not true for \textit{VOBall} and \textit{VOSSIA} data access currently, it is true for the metadata part at least. For the cone services the idea can be pushed also for the data access, since cone services expose catalog data, that is (usually) in tabular format. For SIA this is not possible because SIA version 2.0 is a dataset discovery protocol, thus dealing only with metadata representations and linking to the actual datasets trough identifiers and URLs.

However, given the special role of metadata container, the TAP implementation in the publishing scenario described in this contribution will not be split in an interface part and an access part (that would probably have meant having the request validation and ADQL query parsing at web frontend and query job management at the backend).

A different solution has been chosen, that will be described in Sec.~\ref{subsubsec:taplib}, using an existing TAP library from an external developer and eventually adapting only the needed parts to our current scenario.

Also, since the metadata content and management of the TAP\_SCHEMA-ta that we use behind the scenes is one critical part of the system and requires many insertion, validation and update tasks, this is currently the only part of the infrastructure that has been provided with a mature administration tool, a TAP\_SCHEMA manager (TASMAN, see Sec.~\ref{subsubsec:tasman}).

\subsubsection{The use of TAPlib}
\label{subsubsec:taplib}

Using the TAPlib\footnote{\url{http://cdsportal.u-strasbg.fr/taptuto/}} library it is possible to set up a working TAP service on top of an existing RDB working out some simple configuration steps. For this reason, instead of trying to develop a new tool to deploy TAP services, we chose to use it.

What we actually checked was the ability of TAPlib to cope with both TAP-1.0 (current) recommendation and the ongoing TAP-1.1 proposed one, especially in terms of fields data typing. That's because the current VO efforts goes towards homogenizing a sort-of data definition language based on the \textit{datatype--arraysize--xtype} triplet following the VOTable [\citenum{2013ivoa.spec.0920O}] and DALI (Data Access Layer Interface, [\citenum{2017ivoa.spec.0517D}]) recommendation contents. Such a goal fits quite well with an homogeneous metadata layer as the one sought by the publishing framework expressed in this proceeding.

Apart from that, TAPlib seems to cover most of our needs, like running multiple TAP instances side by side, configuring schemata or table name mapping, plus it is capable of working on top of a PgSphere plugin benefiting from the spherical geometry support. Things we will probably extend will be: the logging interface and the VOSI (Virtual Observatory Support Interfaces, [\citenum{2017ivoa.spec.0524G}]) \textit{/availability} resource, to uniform them into proper auxiliary modules common to all of the discovery and access modules of this publishing toolkit effort (see Sec.~\ref{sec:other}).

\subsubsection{TASMAN TAP\_SCHEMA Manager}
\label{subsubsec:tasman}

As hinted before, due to the fact that the metadata content of TAP\_SCHEMA-ta play a central role in the architecture, some effort has been spent in creating a useful administration tool for this set of relational tables. This led to TASMAN\footnote{\url{http://ia2.inaf.it/index.php/13-software/36-tap-schema-manager}}.

\begin{figure} [ht]
   \begin{center}
   \begin{tabular}{c} %% tabular useful for creating an array of images 
   \includegraphics[width=\textwidth]{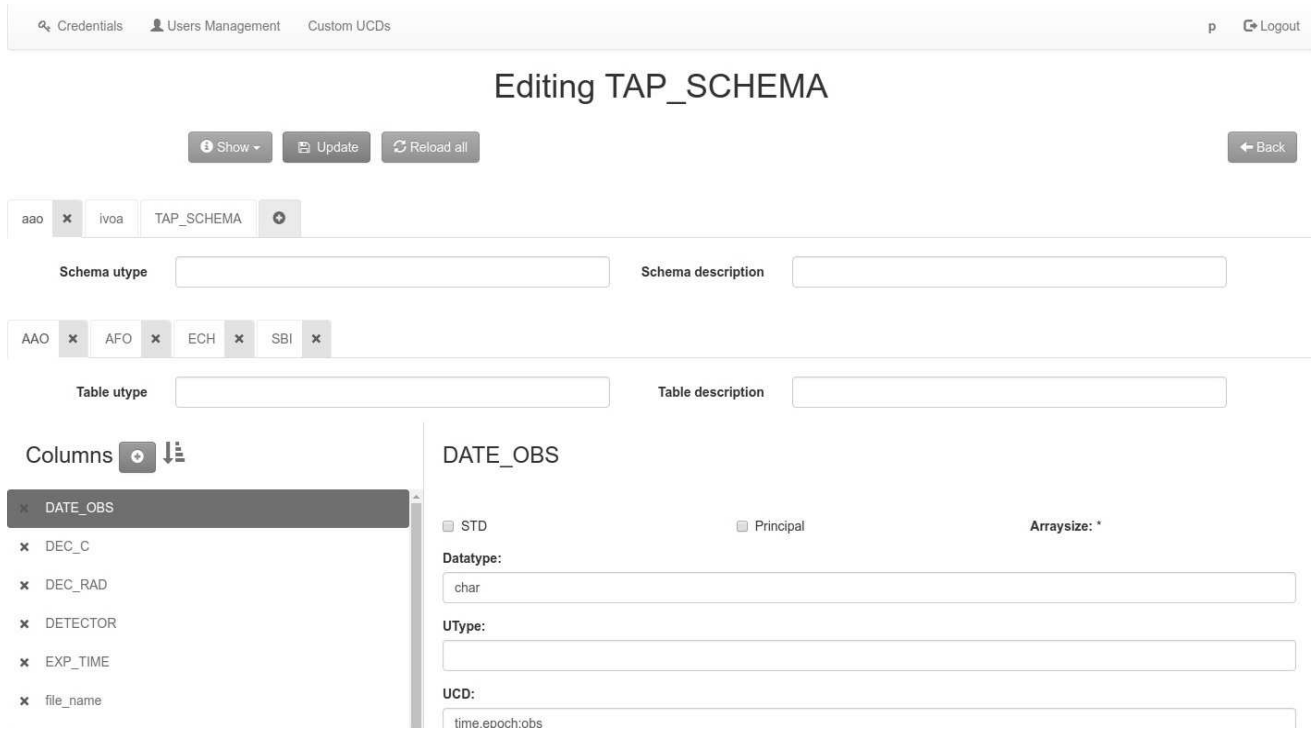}
   \end{tabular}
   \end{center}
   \caption[tasman-screenshot] 
%>>>> use \label inside caption to get Fig. number with \ref{}
   {\label{fig:tasman-screenshot} TASMAN GUI}
   \end{figure} 
   
TASMAN (TAp\_Schema MANager) is a Java EE application that provides a graphical user interface that can be used to create or edit a TAP\_SCHEMA also by people that don't have specific SQL skills. It can be used to edit the ObsCore tables too. Currently it supports both MySQL or PostgreSQL database back-ends.

TASMAN is able to work with multiple TAP\_SCHEMA versions, moreover it can be recompiled for supporting also customized TAP\_SCHEMA structures (e.g. for adding non-standard columns).
It includes also utilities for validating UCD (Unified Content Descriptors, [\citenum{2007ivoa.spec.0402M}], using Ucidy library\footnote{\url{https://github.com/gmantele/ucidy}}), validating VOUnits ([\citenum{2014ivoa.spec.0523D}], using Unity library\footnote{\url{https://bitbucket.org/nxg/unity}}) and generating column index values positioning columns using drag and drop. This latter feature follows the TAP-1.1 ongoing IVOA Proposed Recommendation, like the data typing solution described in the previous (\ref{subsubsec:taplib}) section.

If the database structure is changed from last edit, TASMAN will also try to detect and fix possible inconsistencies. It also allows to store temporary/custom UCDs to allow prototyping of new words in the vocabulary.

TASMAN can be used as an embedded application, i.e. embedding a Jetty web container, or it can be deployed on a server like GlassFish or Tomcat. In both cases it simply needs a browser to be used. It is possible to create multiple TASMAN users and store database credentials and TAP\_SCHEMA settings used by each user.

Even if initially thought to let non-SQL experts to work on TAP\_SCHEMA instances, it is also a valuable tool for expert TAP users, relieving them from typing large amounts of SQL statements to update fields descriptions, annotations and other metadata details, letting the focus be on the metadata enrichment and resource management.

The ability to identify, create and partially validate (as table structure, not the content records' values) the ObsCore standard table for observational datasets, allows the user to take care in an easier way also of that task checking for correct field naming and attributes.

\section{Custom solutions: the VLKB example}
\label{sec:vlkb}

Section~\ref{sec:voimpl} and subsections described the modular architecture in terms of the implementation of VO standards. However it was also said that this solution would prove valuable in creating custom solution, e.g. based upon data center requirements.

To illustrate this we here describe briefly the use case of the VIALACTEA\footnote{\url{http://vialactea.iaps.inaf.it}} Knowledge Base (VLKB, see also [\citenum{2016SPIE.9913E..0HM}]), that required some specific tasks, not yet standardized by IVOA at implementation time or difficult to manage in a VO scenario in the limited time available to the project (being VIALACTEA an EU-FP7 funded program).

The VLKB consists of a set of discovery and access services, deploying a heterogeneous collection of source catalogs (compact and diffuse), images and radio observational data cubes, allowing positional search combined to spectral axes filtering and other constraints and direct cut or mosaicking of the original datasets.

It is composed of a TAP service providing access to the catalog part of the database, a discovery interface to identify the wanted cubes and images, and two services devoted to cutting the images or cubes based on positional or spectral axis bounds and merging adjacent datasets whether the positional boundaries overlap multiple of them.

The set of services, even if they were developed keeping in mind the VO scenario and the modular solution, were initially deployed as a simple set of application insisting on the database structures and collections of datasets.

The underlying libraries for the cutout and the mosaicking part required however mixing the Java language on the client-faced interfaces and C/C++ applications for the actual data manipulation tasks. This was initially achieved using a direct JNI solution, however it proved to be complicated and difficult to scale horizontally.

What happened then was that the connection among the request accepting modules (all the three of them: discovery, cut and merge) was moved into the AMQP messaging solution. This allowed for a better system in the flow of the search, having independent modules take care of independent tasks, but later also allowed for easier updates to the discovery interface, that is now able to provide results mixing full dataset retrieval (where contraints overlap metadata content) and cutout or mosaicking predefined calls based on the constraints used in the discovery phase.

Moreover, while the catalogs discovery and access was since the beginning charged to the TAP service, the metadata of the datasets discovery were only stored in the same database and exposed through TAP, but not used in an actual VO fashion (due to the fact that everything is mainly based on the galactic coordinates frame while IVOA usually mandates the equatorial ICRS frame). Currently a change is ongoing to move towards an ObsCore solution also for this module, thus re-using elements from the common architecture.

Thus, even if it took some time and rework, this quite complex example for the VLKB proved that co-existence of VO and custom modules works and allows for an homogeneous layer for the provided data resources.

\section{Other operational and dedicated modules}
\label{sec:other}

In all of the above the modules and tasks description focused on the main service blocks: discovery interfaces, access modules, the message brokering system. However additional blocks are in place, or will be in place in the future. Here we describe some, being aware that a modular architecture can allow for more, possibly custom, tasks.

\subsection{Logging}
One of these, already cited, is the logging module, to take care of all the query profiling and error logging, so to have a way to describe actual service usage rates, common errors and hints on possible improvements. The logging module is currently really simple and not yet totally homogeneous among the cone search, SIA, TAP and VLKB parts, but updates are foreseen.

For this we tested also Graylog\footnote{\url{https://www.graylog.org/}}, an open source log manager application. Graylog acts as a centralized logs storage, which can collect logs from a very wide set of data sources, process them and even trigger some action when particular situations happen (e.g. sending an alert email message if an error log is detected).

One of the simplest way for sending a log message to Graylog is to use Linux netcat command for creating a TCP or UDP packet containing JSON data. However, Graylog is able to use also more complex systems, like AMQP brokers, or it can be integrated inside a Java application with a special Log4j appender. This means it would be fit in our message driven scenario.

It provides also a graphical administration panel from which it is possible to configure some dashboards for displaying statistics computed from the logs. One interesting option is the possibility to perform geolocation of the data and then anonymize original IP addresses extracted from the logs.

\subsection{VOSI resources}
Other modules that may come in handy are those managing some aspects of the IVOA VOSI interfaces: \textit{/capabilities}, \textit{/availability} and \textit{/examples} (while we leave out the \textit{/tables} resource that is directly connected to the underlying table sets of the data resource).

Capabilities and examples are probably better considered as part of the resource configuration, thus their part will eventually be taken care by the ReST interface in connection with the configuration repository module. This means it will require some work effort on the curation of the resources.

Availability, on the opposite, is part of the status of the resource and, as a consequence it can be considered as a separate task, checking (periodically or on client request) whether a resource interface is active or not and for what reason. Ideally this would be part of a monitoring system, possibly with cascading monitoring and checks. What we are thinking, currently, is to connect it to the Nagios\footnote{\url{https://www.nagios.org/}} system that is already in place at the IA2 data center.

\subsection{VOTable serializer}
Responses from current VO services are serialized in the common VOTable standard. This operation is taken care of by the data access itself (or the TAP module directly). However, VOTable is a data format that includes its own tabular model with elements that allow for proper fields annotation and referencing, but it can be used also to distribute data using specific models, like the really simple \textit{identifier+position} provided by a Simple Cone Search or the ObsCore-like model included in the SIA-2.0 protocol and so on.

For this reason a VOTable dedicated module, able to take as an input a tabular result-set plus a model identifier (or instance) and provide, as an output, the desired VOTable, would be quite useful.

It can be used also for VOTables with embedded Datalink resource descriptors.

\subsection{Resource Document generation}
Another useful module would be the one that, given a data collection and its attached interfaces, could provide the XML document representing it as a VO Resource in terms of the VO Registry (see, as a starting point, [\citenum{2008ivoa.spec.0222P}]).

Even if it could be actually a task that can, again, be part of the configuration repository application, this one can become fairly complex, plus it could be used to directly feed a Resource Registry in itself.

\section{Advantages and drawbacks}
\label{sec:procon}

The modular approach for a resource publishing framework described in this proceeding has, of course, advantages and disadvantages. Out of the first implementation blocks developed within this architecture we can already give some hints on both of them.

First, quite trivial, advantage of modularity is the ability to swap modules with new ones for the same task without the need to touch the full publishing structure. Another is the chance of re-using easily third party libraries encapsulating them in singleton modules and not integrating them in complex systems.

Also the possibility to use the broker system to distribute the load, including redundancy, is an advantage, like is the chance to use different coding languages (provided they have, in our case, an AMQP implementation).

The usage of the TAP\_SCHEMA metadata layer and TAP services providing a common ADQL query language can help also connecting other than API services, e.g. web portals, like IA2 does using its APOGEO\footnote{\url{http://ia2.inaf.it/index.php/13-software/35-apogeo}} (Automated POrtal GEneratOr) application.

As for the disadvantages, it can easily be stated that the maintenance scenario for such an architecture may grow a lot more complex than the one for a monolithic approach, and not necessarily balanced by a smaller coding effort for new functionalities.

Probably, however, the current big disadvantage lives in the internal interfacing among the various modules. This interface is currently custom, but, to have a collaborative grow of the architecture (i.e. outside the scope of one data center only) this has to be changed, or at least standardized enough to be re-used by others. In a sense this architecture describes workflows of query jobs, thus a workflow description language may fit this scope.

\section{Conclusions}
\label{sec:conclusions}

We described here the first implementation steps in a publishing framework inspired by VO recommendations that keeps an eye also on custom requirements. The approach is module and message driven and allows for a distributed environment. It tries also to re-use existing tools and libraries leveraging an integration effort versus a full-coding one.

The collection of modules is currently incomplete and under development, however it already shows some advantages and its re-usability. Besides modules that are already foreseen to complete the framework probably a registry counterpart is a main missing block, while, with respect to the generalization of this effort, a better described or standardized messaging interface among the various modules may be needed. The solution of a message interface considered as a part of a request/response workflow would probably lead towards some workflow language, like the CWL (Common Workflow Language, [\citenum{Amstutz2016}]).

One other effort that would probably give an added value to the framework would be having a global TAP service on top of the multiple single TAPs, one of each we could imagine on top of one or one group of data resources. This could provide a nice global view over the full data center implemeting this modular framework (efforts already exist outside IA2 for such an abstract TAP layer, see, e.g. FireThorn\footnote{\url{http://wfau.metagrid.co.uk/site/firethorn/}}).

\acknowledgments % equivalent to \section*{ACKNOWLEDGMENTS}       
MM likes to acknowledge Francesco Cepparo for its degree thesis work leading to the message-driven scenario for a distributed publishing solution. The thesis was a joint effort of the IA2 data center, part of INAF, and the Department of Mathematics and Informatics of the University of Udine. MM likes also to acknowledge all the software development made by its co-authors under the funding of RS, the IA2 project lead.

% References
\bibliography{10707-74} % bibliography data in report.bib

\begin{thebibliography}{10}

\bibitem{2008ivoa.specQ0222P}
{Plante}, R., {Williams}, R., {Hanisch}, R., {Szalay}, A., and {Plante}, R.,
  ``{Simple Cone Search Version 1.03},'' tech. rep. (Feb. 2008).

\bibitem{2009ivoa.spec.1111H}
{Harrison}, P., {Tody}, D., {Plante}, R., and {Harrison}, P., ``{Simple Image
  Access Specification Version 1.0},'' tech. rep. (Nov. 2009).

\bibitem{2012ivoa.spec.0210T}
{Tody}, D., {Dolensky}, M., {McDowell}, J., {Bonnarel}, F., {Budavari}, T.,
  {Busko}, I., {Micol}, A., {Osuna}, P., {Salgado}, J., {Skoda}, P.,
  {Thompson}, R., {Valdes}, F., {Data Access Layer Working Group}, and {Tody},
  D., ``{Simple Spectral Access Protocol Version 1.1},'' tech. rep. (Feb.
  2012).

\bibitem{2012SPIE.8451E..05M}
{Molinaro}, M., {Knapic}, C., and {Smareglia}, R., ``{The VO-Dance web
  application at the IA2 data center},'' in [{\em Software and
  Cyberinfrastructure for Astronomy II. Proceedings of the SPIE, Volume 8451,
  article id. 845105, 12 pp. (2012).}{\nolinebreak\hspace{0.1em}]},   {\bf
  8451} (Sept. 2012).

\bibitem{2012ASPC..461..419M}
{Molinaro}, M., {Laurino}, O., and {Smareglia}, R., ``{Integrating the IA2
  Astronomical Archive in the VO: The VO-Dance Engine},'' in [{\em Astronomical
  Data Analysis Software and Systems XXI. Proceedings of a Conference held at
  Marriott Rive Gauche Conference Center, Paris, France, 6-10 November, 2011.
  ASP Conference Series, Vol. 461. Edited by P. Ballester, D. Egret, and N.P.F.
  Lorente. San Francisco: Astronomical Society of the Pacific, 2012.,
  p.419}{\nolinebreak\hspace{0.1em}]},   {\bf 461},  419 (Sept. 2012).

\bibitem{2010ivoa.spec.0327D}
{Dowler}, P., {Rixon}, G., {Tody}, D., and {Dowler}, P., ``{Table Access
  Protocol Version 1.0},'' tech. rep. (Mar. 2010).

\bibitem{2014SPIE.9152E..0CM}
{Molinaro}, M., {Cepparo}, F., {De Marco}, M., {Knapic}, C., {Apollo}, P., and
  {Smareglia}, R., ``{Modular VO oriented Java EE service deployer},'' in [{\em
  Proceedings of the SPIE, Volume 9152, id. 91520C 8 pp.
  (2014).}{\nolinebreak\hspace{0.1em}]},   {\bf 9152} (July 2014).

\bibitem{2016SPIE.9913E..28C}
{Cepparo}, F., {Scagnetto}, I., {Molinaro}, M., and {Smareglia}, R., ``{A
  distributed infrastructure for publishing VO services: an implementation},''
  in [{\em Proceedings of the SPIE, Volume 9913, id. 991328 6 pp.
  (2016).}{\nolinebreak\hspace{0.1em}]},   {\bf 9913} (July 2016).

\bibitem{2014ASPC..485..139M}
{Molinaro}, M., {Apollo}, P., {Knapic}, C., and {Smareglia}, R., ``{Building an
  Archive Backbone Extending TAP},'' in [{\em Astronomical Data Analysis
  Software and Systems XXIII. Proceedings of a meeting held 29 September - 3
  October 2013 at Waikoloa Beach Marriott, Hawaii, USA. Edited by N. Manset and
  P. Forshay ASP conference series, vol. 485, 2014,
  p.139}{\nolinebreak\hspace{0.1em}]},   {\bf 485},  139 (May 2014).

\bibitem{2016SPIE.9913E..0HM}
{Molinaro}, M., {Butora}, R., {Bandieramonte}, M., {Becciani}, U., {Brescia},
  M., {Cavuoti}, S., {Costa}, A., {Di Giorgio}, A.~M., {Elia}, D., {Hajnal},
  A., {Gabor}, H., {Kacsuk}, P., {Liu}, S.~J., {Molinari}, S., {Riccio}, G.,
  {Schisano}, E., {Sciacca}, E., {Smareglia}, R., and {Vitello}, F.,
  ``{VIALACTEA knowledge base homogenizing access to Milky Way data},'' in
  [{\em Proceedings of the SPIE, Volume 9913, id. 99130H 11 pp.
  (2016).}{\nolinebreak\hspace{0.1em}]},   {\bf 9913} (Aug. 2016).

\bibitem{2010ivoa.rept.1123A}
{Arviset}, C., {Gaudet}, S., {IVOA Technical Coordination Group}, and
  {Arviset}, C., ``{IVOA Architecture Version 1.0},'' tech. rep. (Nov. 2010).

\bibitem{2017ivoa.spec.0509L}
{Louys}, M., {Tody}, D., {Dowler}, P., {Durand}, D., {Michel}, L., {Bonnarel},
  F., {Micol}, A., {IVOA DataModel Working Group}, {Louys}, M., {Tody}, D.,
  {Dowler}, P., and {Durand}, D., ``{Observation Data Model Core Components,
  its Implementation in the Table Access Protocol Version 1.1},'' tech. rep.
  (May 2017).

\bibitem{2015ivoa.spec.1223D}
{Dowler}, P., {Bonnarel}, F., {Tody}, D., {Dowler}, P., and {Bonnarel}, F.,
  ``{IVOA Simple Image Access Version 2.0},'' tech. rep. (Dec. 2015).

\bibitem{2017ivoa.spec.0517B}
{Bonnarel}, F., {Dowler}, P., {Demleitner}, M., {Tody}, Douglas, D.~J., and
  {Fran{\c{c}}ois Bonnarel}, P.~D., ``{IVOA Server-side Operations for Data
  Access Version 1.0},'' tech. rep. (May 2017).

\bibitem{2015ivoa.spec.0617D}
{Dowler}, P., {Bonnarel}, F., {Michel}, L., {Demleitner}, M., and {Dowler}, P.,
  ``{IVOA DataLink Version 1.0},'' tech. rep. (June 2015).

\bibitem{2008ivoa.spec.1030O}
{Osuna}, P., {Ortiz}, I., {Lusted}, J., {Dowler}, P., {Szalay}, A.,
  {Shirasaki}, Y., {Nieto- Santisteban}, M.~A., {Ohishi}, M., {O'Mullane}, W.,
  {VOQL-TEG Group}, {VOQL Working Group.}, {Osuna}, P., and {Ortiz}, I.,
  ``{IVOA Astronomical Data Query Language Version 2.00},'' tech. rep. (Oct.
  2008).

\bibitem{2013ivoa.spec.0920O}
{Ochsenbein}, F., {Taylor}, M., {Williams}, R., {Davenhall}, C., {Demleitner},
  M., {Durand}, D., {Fernique}, P., {Giaretta}, D., {Hanisch}, R., {McGlynn},
  T., {Szalay}, A., {Wicenec}, A., {Ochsenbein}, F., and {Taylor}, M.,
  ``{VOTable Format Definition Version 1.3},'' tech. rep. (Sept. 2013).

\bibitem{2017ivoa.spec.0517D}
{Dowler}, P., {Demleitner}, M., {Taylor}, M., {Tody}, D., and {Dowler}, P.,
  ``{Data Access Layer Interface Version 1.1},'' tech. rep. (May 2017).

\bibitem{2017ivoa.spec.0524G}
{Graham}, M., {Rixon}, G., {Dowler}, P., {Major}, B., {Grid}, {Web Services
  Working Group}, {Graham}, M., {Rixon}, G., {Dowler}, P., and {Major}, B.,
  ``{IVOA Support Interfaces Version 1.1},'' tech. rep. (May 2017).

\bibitem{2007ivoa.spec.0402M}
{Preite Martinez}, A., {Derriere}, S., {Delmotte}, N., {Gray}, N., {Mann}, R.,
  {McDowell}, J., {Mc Glynn}, T., {Ochsenbein}, F., {Osuna}, P., {Rixon}, G.,
  {Williams}, R., {Preite Martinez}, A., and {Derriere}, S., ``{The UCD1+
  controlled vocabulary Version 1.23},'' tech. rep. (Apr. 2007).

\bibitem{2014ivoa.spec.0523D}
{Derriere}, S., {Gray}, N., {Demleitner}, M., {Louys}, M., {Ochsenbein}, F.,
  {Derriere}, S., and {Gray}, N., ``{Units in the VO Version 1.0},'' tech. rep.
  (May 2014).

\bibitem{2008ivoa.spec.0222P}
{Plante}, R., {Benson}, K., {Graham}, M., {Greene}, G., {Harrison}, P.,
  {Lemson}, G., {Linde}, T., {Rixon}, G., {St{\'e}b{\'e}}, A., {IVOA Registry
  Working Group}, and {Plante}, R., ``{VOResource: an XML Encoding Schema for
  Resource Metadata Version 1.03},'' tech. rep. (Feb. 2008).

\bibitem{Amstutz2016}
Amstutz, P., Crusoe, M.~R., Tijanić, N., Chapman, B., Chilton, J., Heuer, M.,
  Kartashov, A., Leehr, D., Ménager, H., Nedeljkovich, M., Scales, M.,
  Soiland-Reyes, S., and Stojanovic, L., ``{Common Workflow Language, v1.0},''
  (7 2016).

\end{thebibliography}
\bibliographystyle{spiebib} % makes bibtex use spiebib.bst

\end{document}